\documentclass{aastex}
\usepackage{emulateapj5}
\input{epsf}

\usepackage{times}

\newcommand{\nh}{N_{\rm H}}
\newcommand{\wfe}{W_{{\rm K}\alpha}}
\newcommand{\ee}{$e^\pm$}
\newcommand{\g}{$\gamma$}
\newcommand{\lh}{\ell_{\rm h}}
\newcommand{\ls}{\ell_{\rm s}}
\newcommand{\lth}{\ell_{\rm th}}
\newcommand{\lnth}{\ell_{\rm nth}}

\newcommand{\source}{GRS 1915+105}

\newcommand{\xte}{{\it RXTE}}

\newcommand{\gro}{{\it CGRO}}
\newcommand{\asm}{{\it RXTE}/ASM}

\begin{document}

\slugcomment{The Astrophysical Journal Letters, in press}

\shorttitle{OSSE AND \xte\/ OBSERVATIONS OF GRS 1915+105}
\shortauthors{ZDZIARSKI ET AL.}

\title{OSSE and \xte\/ Observations of GRS 1915+105: \\
Evidence for Non-thermal Comptonization}

\author{Andrzej A. Zdziarski,\altaffilmark{1}
J. Eric Grove,\altaffilmark{2}
Juri Poutanen,\altaffilmark{3}
A. R. Rao,\altaffilmark{4}
and S. V. Vadawale\altaffilmark{4}
}
\altaffiltext{1}{N. Copernicus Astronomical Center, Bartycka 18,
00-716 Warsaw, Poland; aaz@camk.edu.pl}

\altaffiltext{2}{E. O. Hulburt Center for Space Research,
Naval Research Laboratory, Washington, DC 20375, USA}

\altaffiltext{3}{Stockholm  Observatory, SE-133 36 Saltsj\"obaden, Sweden}

\altaffiltext{4}{Tata Institute of Fundamental Research, Homi Bhabha Road,
Bombay 400 005, India}


\begin{abstract}
GRS 1915+105 was observed by the {\it CGRO}/OSSE 9 times in 1995-2000, and 8 of
those observations were simultaneous with those by \xte. We present an analysis
of all of the OSSE data and of two \xte-OSSE spectra with the lowest and
highest X-ray fluxes. The OSSE data show a power-law--like spectrum extending
up to $\ga 600$ keV without any break. We interpret this emission as
strong evidence for the presence of non-thermal  electrons in the source.
The broad-band spectra cannot be described by either thermal or bulk-motion
Comptonization, whereas they are well described by Comptonization in hybrid
thermal/non-thermal plasmas.
\end{abstract}

\section {Introduction}
\label{intro}

The black-hole binary \source\ is highly variable in X-rays (Belloni et
al.\ 2000, and references therein). Still, even its hardest spectra are
relatively soft, consisting of a blackbody-like component and a high-energy
tail (Vilhu et al.\ 2001). They are softer than those of other black-hole
binaries in the hard state, which $EF_E$ spectra peak at $\sim 100$ keV (e.g.,
Cyg X-1, Gierli\'nski et al.\ 1997), and are similar to their soft state (e.g.,
Cyg X-1, Gierli\'nski et al.\ 1999, hereafter G99; LMC X-1, LMC X-3, Wilms et 
al.\ 2001).

The blackbody component arises, most likely, in an optically-thick accretion
disk. On the other hand, there is no consensus at present regarding the origin
of the tail. All three main models proposed so far involve Comptonization of
the blackbody photons by high-energy electrons. They differ, however, in the
distribution (and location) of the electrons, which are assumed to be either
thermal (Maxwellian), non-thermal (close to a power law), or in a free fall
onto the black hole.

A discussion of these models is given in Zdziarski (2000), who shows that the
thermal and free-fall models of the soft state of black hole binaries can be
ruled out, mostly by the marked absence of a high-energy cutoff around 100 keV
in the \gro\/ data (Grove et al.\ 1998; G99; Tomsick et al.\ 1999; McConnell et
al.\ 2000). The present best soft-state model appears to involve electron
acceleration out of a Maxwellian distribution (i.e., a non-thermal process),
which leads to a hybrid electron distribution consisting of both thermal and
non-thermal parts (Zdziarski, Lightman \& Macio{\l}ek-Nied\'zwiecki 1993;
Poutanen \& Coppi 1998; G99; Coppi 1999).

In this {\it Letter}, we present all OSSE observations of \source. We then
choose two OSSE spectra corresponding to the lowest and highest X-ray flux and
fit them together with spectra from simultaneous  \xte\/ pointed observations.
The spectra, showing extended power laws without any cutoff up to at least 600
keV, provide strong evidence for the presence of non-thermal Comptonization.
More extensive presentation of the combined  X-ray/OSSE data will be given
elsewhere.

\section {OSSE and \xte\/ spectra} \label{osse}

Table 1 gives the log of the 9 OSSE observations, together with results of
power-law fits and basic data about the corresponding X-ray and radio states.
The OSSE instrument accumulated spectra in a sequence of 2-min.\ measurements
of the source field alternated with 2-min., offset-pointed measurements of
background.  The background spectrum for each source field was derived
bin-by-bin with a quadratic interpolation in time of the nearest background
fields (see Johnson et al.\ 1993). Figure \ref{fig:osse} shows the OSSE spectra
(including standard energy-dependent systematic errors), which were fitted up
to energies at which the source signal was still detected. The uncertainty for
a fitted parameter corresponds hereafter to 90\% confidence ($\Delta \chi^2 =
2.71$). We see that the source went through wide ranges of radio and X-ray
fluxes and types of X-ray variability during those observations. In spite of
that variety, 8 out of 9 OSSE spectra are best-fitted by a power  law with a
photon index of $\Gamma\simeq 3.0\pm 0.1$ and the flux varying within a  factor
of 2. The only exception is the OSSE spectrum corresponding to the  highest
X-ray flux measured by the ASM (1999 April 21--27), which is much  harder,
$\Gamma\simeq 2.3$, and has a much lower flux.

We then consider the OSSE spectra corresponding to the extreme X-ray fluxes
measured by the \asm, i.e., from 1997 May 14--20 (VP 619) and 1999 April\
21--27 (VP 813). We fit them together with spectra from the pointed \xte\/
observations of 1997 May 15 and 1999 April 23 (the observation IDs are
20187-02-02-00, 40403-01-07-00; 1\% systematic error is added to the PCA data
with the responses of 2001 February).  These PCA data correspond to the
variability classes (Belloni et al.\ 2000) of $\chi$ and $\gamma$, in which the
variability is moderate and the source spends  most of the time in two basic
low ($C$) and high ($B$) X-ray flux state, respectively.

We fit the data with the XSPEC (Arnaud 1996) model {\tt eqpair} (Coppi 1999;
G99), which calculates self-consistently microscopic processes in a hot plasma
with electron acceleration at a power law rate with an index, $\Gamma_{\rm
inj}$, in a background thermal plasma with a Thomson optical depth of
ionization electrons, $\tau_{\rm i}$. The  electron temperature, $kT$, is
calculated from the balance of Compton and  Coulomb energy exchange, as well as
\ee\ pair production (yielding the total optical depth of $\tau>\tau_{\rm i}$)
is taken into account.  The last two processes depend on the plasma
compactness, $\ell\equiv {\cal  L}\sigma_{\rm T}/({\cal R} m_{\rm e} c^3)$,
where ${\cal L}$ is a power  supplied to the hot plasma, ${\cal R}$ is its
characteristic size, and  $\sigma_{\rm T}$ is the Thomson cross section. We
then define a hard compactness, $\lh$, corresponding to the power supplied to
the electrons, and a soft compactness, $\ls$, corresponding to the power in
soft seed photons irradiating the plasma (which are assumed to be emitted by a
blackbody disk with the maximum temperature, $kT_{\rm bb}$). The compactnesses
corresponding  to the electron acceleration and to a direct heating (i.e., in
addition to Coulomb energy exchange with non-thermal \ee\ and Compton heating)
of the  thermal \ee\ are denoted as $\lnth$ and $\lth$, respectively, and
$\lh=\lnth +  \lth$. Details of the model are given in G99.

\medskip
\epsscale{1.0} \centerline{\epsfxsize=8.7cm
\epsfbox{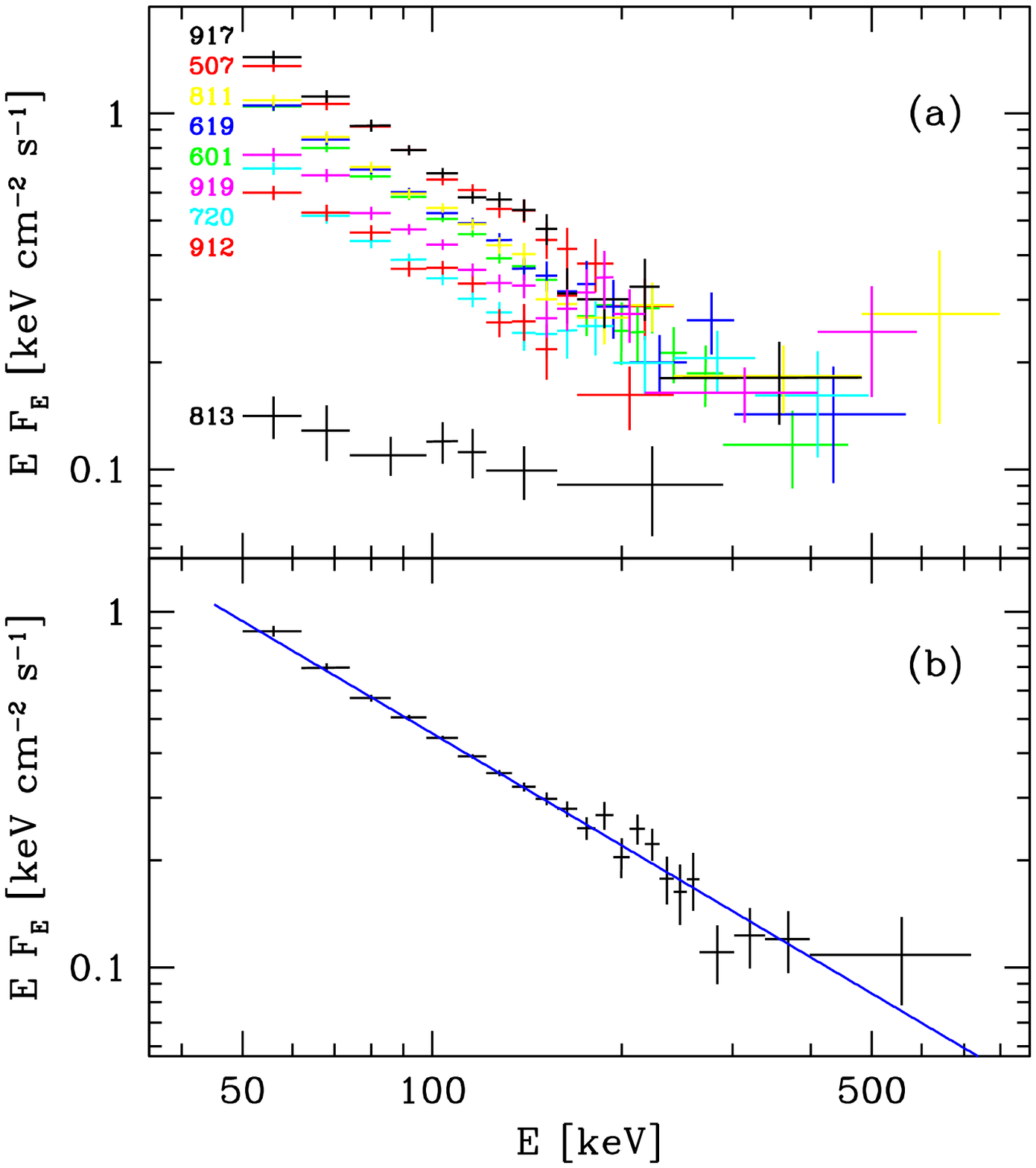}}
\figcaption{(a) The 9 OSSE spectra of \source\ deconvolved assuming power-law
models. The color-coded spectra are labeled by the OSSE observation numbers
(Table 1) ordered according to the 50--60 keV flux. (b) The average spectrum
of all observations with its power-law best fit. When fitted by an e-folded
power law, the lower limit on the e-folding energy is 0.9\,MeV.
\label{fig:osse} }
\medskip

We also take into account Compton reflection with a solid angle of $\Omega$ 
(Magdziarz \& Zdziarski 1995) and an Fe K$\alpha$ emission from an accretion 
disk assumed to extend down to $10GM/c^2$ (which results in a relativistic 
smearing). The equivalent width, $\wfe$, with respect to the {\it scattered\/} 
spectrum only  is tied to $\Omega$ via $\wfe\simeq 100(\Omega/2\pi)$ eV (George 
\& Fabian  1991). The elemental abundances of Anders \& Ebihara (1982), an 
absorbing column of $\nh\geq 1.75\times  10^{22}$ cm$^{-2}$ (Dickey \& Lockman 
1990; Vilhu et al.\ 2001), and an inclination of $i=70\degr$ are assumed.

As discussed in G99, $\chi^2$ depends weakly on $\ls$ in a wide range of this
parameter. An increase of $\ls$ leads to increasing \ee\ pair production, which
then leads to an annihilation feature around 511 keV. The presence of such a
feature is compatible with the OSSE data (Fig.\ \ref{fig:osse}), but only very
weakly constrained. G99 found that $\ls=10$ provides a good fit to Cyg X-1
data. Here, we find that a good fit is provided with $\ls=100$, compatible with
the high luminosity of \source. For example, for 1/2 of the Eddington
luminosity, $L_{\rm E}$, and spherical geometry, the size of the plasma
corresponds then to $\sim 100GM/c^2$.

This model provides very good description of our two broad-band spectra (as
well as of other \xte-OSSE spectra, S. V. Vadawale et al., in preparation). For
VP 619, we assume a free relative normalization of the HEXTE and OSSE spectra
with respect to those from the PCA. On the other hand, the HEXTE spectrum for
VP 813 has relatively few counts at its highest energies, and thus we use the
actual OSSE normalization in that fit. Table 2 gives the fit results, and
Figure \ref{f:data} shows the spectra.

\medskip
\epsscale{1.0}
\centerline{\epsfxsize=8.2cm \epsfbox{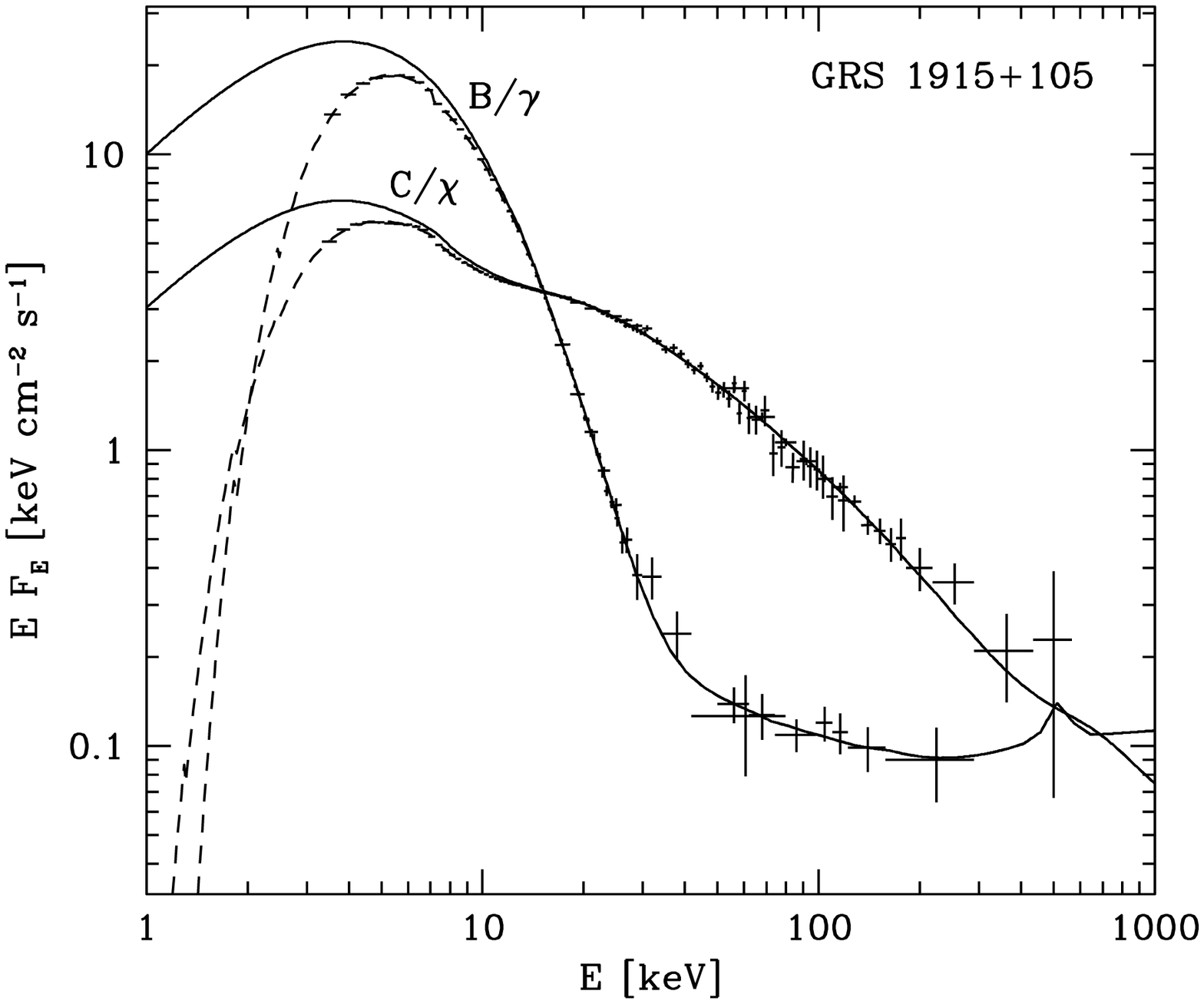}}
\figcaption{Fits to simultaneous PCA-HEXTE-OSSE spectra from VP 619 and 813
with the hybrid Comptonization model. The dashed and solid curves show the
models of the observed spectra and the unabsorbed spectra, respectively. The
data are normalized to the PCA.
\label{f:data}
}
\medskip

Our model predicts the power law emission extending with no cutoff well above 1
MeV and a weak annihilation feature (with the plasma allowed to be
pair-dominated, i.e., with $\tau_{\rm i}\rightarrow 0$,  for VP 619). Those
predictions can be tested by future soft \g-ray detectors more sensitive than
the OSSE. We note that the COMPTEL has already detected a power law tail up to
$\sim$5--10 MeV in the soft state of Cyg X-1 (McConnell et al.\ 2000).

Figure \ref{f:model}a shows the spectral components of the fit to the VP-619
spectrum. Compton reflection with $\Omega\sim 2\pi$ is detected at a very high
significance ($\Omega=0$ gives $\chi^2/\nu=214/131$, corresponding to the
chance appearance of reflection of $6\times 10^{-26}$ from the $F$-test), and
it is responsible for the convex curvature in the $\sim$10--100 keV
spectrum. Figure \ref{f:model}a also shows that the scattered component has a
spectral break at $\sim 100$ keV but continues as a power law (with addition of
the broad annihilation feature) above it due to the domination of non-thermal
scattering at those energies.  Comptonization by the thermal electrons
dominates at energies close to the blackbody  component and thus PCA data of
\source\ can be reasonably modeled up to 60 keV by thermal Comptonization of a
disk blackbody (Vilhu et al.\ 2001). When the OSSE data are included, the
probability that non-thermal electrons are not present (i.e., $\lnth =0$) is
only $4\times 10^{-10}$. The best-fit thermal Compton model shown by the long
dashes in Figure \ref{f:model}a strongly underestimates the flux above 100 keV.
The statistical significance of the presence of non-thermal electrons can be
further increased by fitting the \xte\/ spectrum together with the average
spectrum from OSSE, which has virtually identical shape to that of VP 619, but
much better statistics. Then, allowing for $\lnth>0$ reduces $\chi^2/\nu$ from
$228/139$ to $105/137$, which corresponds to the chance probability of $4\times
10^{-24}$. Thus, we strongly rule out the pure-thermal Comptonization model.

During the VP 813, Compton reflection is statistically not required, as indeed
expected at the large $\tau\sim 4$ of the scattering medium covering the disk
(which would completely smear out any disk reflection and fluorescence
features). Thus, we set $\Omega=0$ is the fit. The presence of non-thermal
electrons is now required at an extremely high significance ($1-10^{-40}$) due
to the presence of the very distinct hard high-energy tail above the
thermally--cut-off spectrum, see Figures \ref{f:data}, \ref{f:model}b.

\medskip
\epsscale{1.0}
\centerline{\epsfxsize=8.4cm \epsfbox{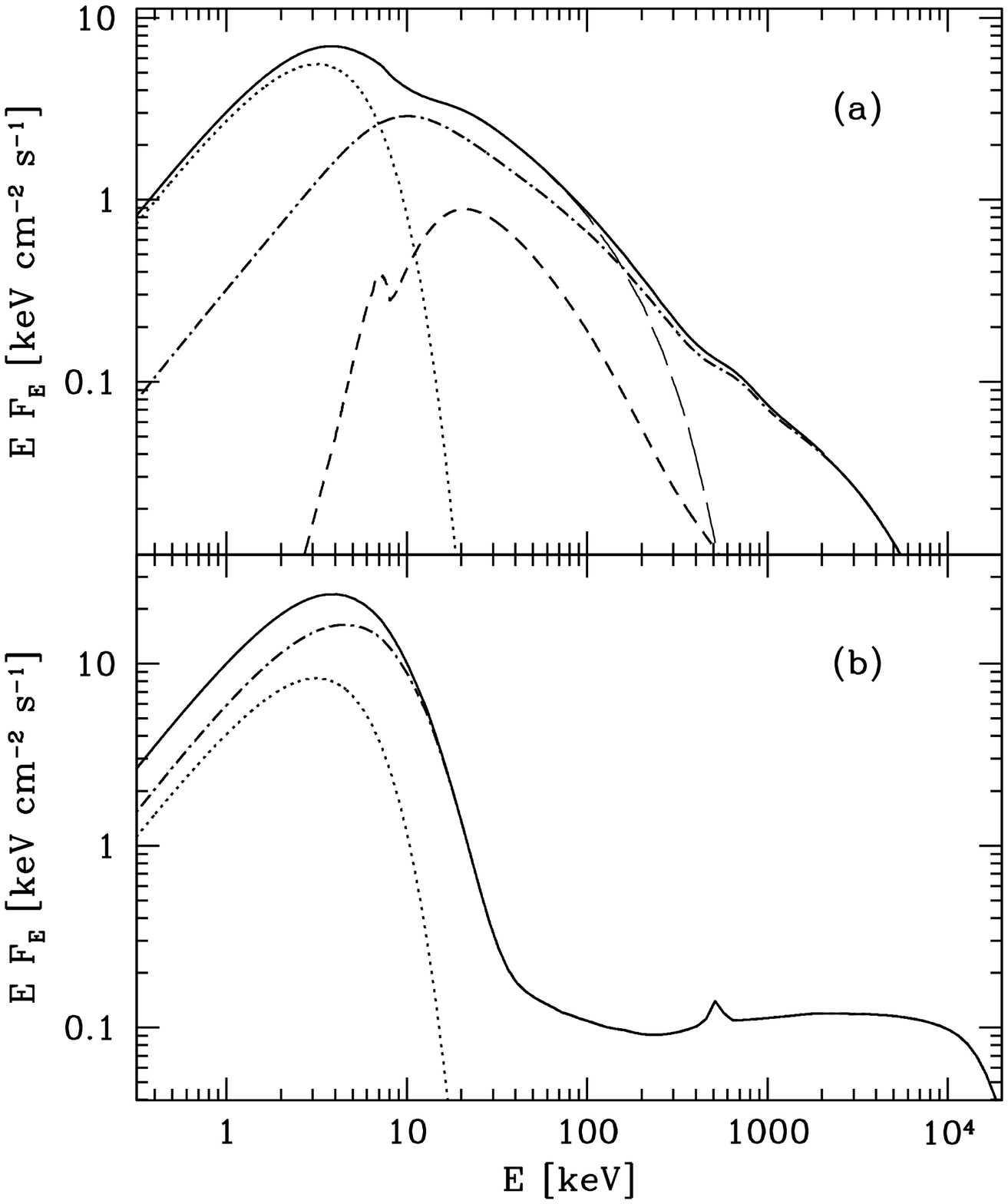}}
\figcaption{(a) Components of the fit to the VP 619 data. All spectra are
intrinsic, i.e., corrected for absorption. The dotted, dot-dashed and dashed
curves give the unscattered blackbody component, the scattered spectrum, and
the component due to Compton reflection and Fe K$\alpha$ fluorescence,
respectively. The solid curve is the total spectrum. The thin long-dashed curve
shows the best-fit thermal Comptonization model, which lies much below the
data above 100 keV. (b) The total model spectrum and the corresponding two
components for the VP 813 data. The cutoff at $\ga 10$ MeV is due to pair
absorption.
\label{f:model}
}
\medskip

On the other hand, Shrader \& Titarchuk (1998) have fitted a model of
bulk-motion Comptonization of blackbody photons to \xte\/ and BATSE data from a
hard state of \source\ similar to that of VP 619. We fit their model ({\tt bmc}
in XSPEC) at a free $N_{\rm H}$ to our broad-band spectra, and find it is
completely unacceptable statistically, with $\chi^2/\nu=851/133$ and $549/92$ for
VP 619 and 813, respectively.

However, the {\it specific\/} feature of bulk-motion Comptonization is a
high-energy cutoff at $\ga 100$ keV due to the effects of Compton recoil and
gravitational redshift close to the black-hole horizon  (e.g.\ Laurent \&
Titarchuk 1999). Such a cutoff is {\it not\/} included in the {\tt bmc} model
(and its inclusion would further worsen the fits above). Thus, we use Monte
Carlo results of Laurent \& Titarchuk (1999), which include the cutoff, to test
whether the OSSE data (regardless of the data at lower energies) are compatible
with its presence. We find that their theoretical spectrum for the accretion
rate of $\dot M= 2L_{\rm E}/c^2$ matches well the slope of the average OSSE
spectrum at low energies (the histogram in Fig.~\ref{f:bmc}). The Monte-Carlo
spectrum can then be very well reproduced by a power law times a step function
convolved with a Gaussian (model {\tt plabs(step)} in XSPEC, with the cutoff
energy of 150 keV and the Gaussian width of 35 keV, the solid curve in
Fig.~\ref{f:bmc}). We see that the OSSE average spectrum lies well above that
model at $\ga 100$ keV.  Quantitatively, the bulk-motion Compton model yields
$\chi^2/\nu= 745/48$. In comparison, the power-law and {\tt eqpair} models
yield $\chi^2/\nu=29/48$ and $\chi^2/\nu=31/44$, respectively. Thus, the
bulk-Compton model is completely ruled out. Further problems with that model
are discussed in Zdziarski (2000).

\medskip
\epsscale{1.0}
\centerline{\epsfxsize=7.0cm \epsfbox{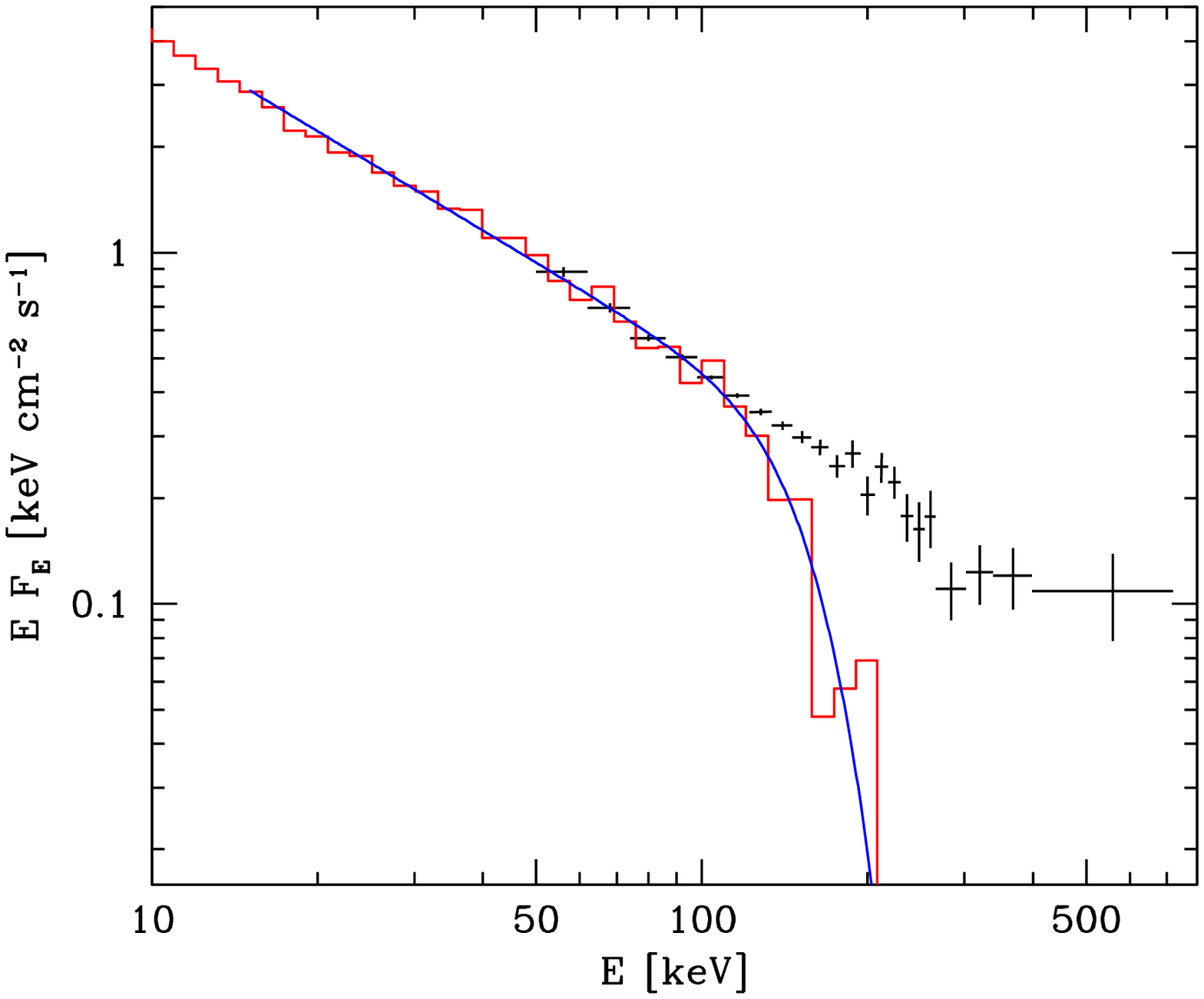}}
\figcaption{Comparison of the average OSSE spectrum fitted by a power law
(crosses) with predictions of the bulk-motion Comptonization model (Laurent \&
Titarchuk 1999; red histogram). The blue solid curve is an analytical approximation of that model, see text. The theoretical spectrum predicts a correct low-energy slope, but it fails to reproduce the data at $\ga 100$ keV.
\label{f:bmc}
}
\medskip

Also, the commonly used phenomenological models of disk blackbody and either a
power law or an e-folded power law give very bad fits to our data. The latter
yields $\chi^2/\nu= 263/132$, $405/91$ for VP 619, 813, respectively. In fact,
even the thermal part of the VP-813 spectrum is very poorly described by a disk
blackbody, with $\chi^2/\nu=3450/36$ for a fit to the 3.5--20 keV PCA data,
whereas the same data are well modeled by thermal Comptonization,
$\chi^2/\nu=22/34$ (with neither model including a high-energy tail). Thus, we
find physical models of the spectra of \source\ in terms of thermal
and non-thermal Comptonization and (in some cases) Compton reflection to be
vastly superior to any other model proposed so far.

\section{Discussion and Conclusions}

We have found that broad-band spectra of \source\ in its two main spectral
states ($B$, $C$) are very well fitted by Comptonization of disk blackbody
photons in a (hybrid) plasma with both electron heating and acceleration. The
presence of strong reflection indicates the plasma is located in coronal
regions (possibly magnetic flares) above the disk. This physical model is the
same as that fitted to the soft state of Cyg X-1 by G99. Differences in the
variability properties of Cyg X-1 and \source\ are likely to be due to the disk
being stable in the former case and unstable in the latter, most likely due to
its much higher $\dot M$. The corona is unstable in both cases. The
Comptonizing medium in the high-$L$ state ($B$) is Thomson-thick, and it can
represent the surface layer of an overheated disk accreting at a
super-Eddington rate (Beloborodov 1998). An issue to be addressed by future
research is the origin of the hardening of the tail in this state as compared
to other ones (Fig.~\ref{fig:osse}). Our model predicts broad annihilation
features, although their strength depends on the unknown size of the plasma. 

\source\ was in the power-law \g-ray state in the classification of Grove et
al.\ (1998) during the 9 OSSE observations. This state usually corresponds to
the high/soft X-ray state. Indeed, X-ray spectra observed so far from \source\
(Vilhu et al.\ 2001) are substantially softer than those with $\Gamma\sim 1.7$
and a sharp thermal cutoff at $E\ga 100$ keV, characteristic to the hard state
of other black-hole binaries.

\acknowledgements

We thank P. Coppi and M. Gierli\'nski for their work on the {\tt eqpair} model,
W. N. Johnson for help with the OSSE data reduction, and Ph.\ Laurent for
supplying his Monte Carlo results. This  research has been  supported by grants
from KBN (2P03D00614 and 2P03C00619p1,2), the Foundation for Polish Science
(AAZ), and the Swedish  Natural  Science  Research  Council and the Anna-Greta
and Holger Crafoord Fund (JP). JP and AAZ acknowledge support from the Royal
Swedish Academy of Sciences, the Polish Academy of Sciences and the Indian
National Science Academy through exchange programs.

\footnotesize
\begin{center}
{\sc TABLE 1\\ OSSE observations of GRS 1915+105}
\vskip 2pt
\begin{tabular}{cllcccccc}
\hline
\hline
&\\
{OSSE} &{Start} &{End}
&{Exposure\tablenotemark{b}} &{$\Gamma$} &{OSSE} &
{X-ray} &{X-ray flux\tablenotemark{e}} &{2.2 GHz flux} \\

{VP\tablenotemark{a}} &{date} &{date} &{[$10^5$ s]}
&{} & {flux\tablenotemark{c}} & {obs.\tablenotemark{d}}
&{[mCrab]} &{[mJy]} \\
&\\
\hline
&\\
507  & 1995 Nov 28 & Dec 7 & 2.26   &   $3.08^{+0.08}_{-0.07}$ &
$18.8^{+0.5}_{-0.4}$ &   & &  \\
601 & 1996 Oct 15 & Oct 29&  7.15   & $3.08^{+0.05}_{-0.05}$  &
$14.1^{+0.2}_{-0.2}$ &
  $\nu \chi \chi \nu \chi $ &  840$\pm$330\tablenotemark{f}  &  \\
 619 & 1997 May 14 & May 20& 3.27    & $3.06^{+0.06}_{-0.06}$
& $14.7^{+0.3}_{-0.4}$  &
 $\alpha  \chi \chi \alpha \chi $ &  270$\pm$120  & 33$\pm$12\tablenotemark{f}
\\
 720 & 1998 May  5 & May 15& 4.20    & $3.00^{+0.09}_{-0.09}$  &
$9.5^{+0.3}_{-0.2}$ &
 $\chi    \chi \chi \chi   \chi$ & 570$\pm$40 &  91$\pm$22  \\
 811 & 1999 Apr 6 & Apr 13&  2.92   & $3.10^{+0.07}_{-0.07}$
&  $14.9^{+0.3}_{-0.4}$ &
 $\alpha \rho    \rho $           & 830$\pm$350 & 11$\pm$6  \\
812 &  1999 Apr 14 & Apr 20& 2.85   & $2.89^{+0.10}_{-0.10}$
&  $9.5^{+0.4}_{-0.3}$ & $\kappa \kappa \kappa \kappa \kappa \kappa $
&920$\pm$270 & 9$\pm$4\\
813 &  1999 Apr 21 & Apr 27& 3.23   & $2.33^{+0.28}_{-0.27}$ &
$3.0^{+0.4}_{-0.3}$ &
   $\gamma$      &1190$\pm$240 & 10$\pm$4\\
917 &  2000 Apr 18 & Apr 25& 1.56  & $3.16^{+0.08}_{-0.07}$
& $19.1^{+0.4}_{-0.4}$ &
 $\chi \chi   \chi \chi \alpha  \alpha  $ & 430$\pm$100 & 19$\pm$7\\
919 & 2000 May 9 & May 26& 3.89  & $2.96^{+0.08}_{-0.08}$
& $11.5^{+0.3}_{-0.3}$ &
$\chi \rho \rho \rho \rho \rho \rho \rho \rho $ & 700$\pm$90 &
\\
Sum &  & & 29.60 & $3.05^{+0.04}_{-0.04}$ & $12.2^{+0.1}_{-0.2}$ \\
\hline
\end{tabular}
\label{t:log}
\end{center}
\setcounter{table}{1}
{
$^{\rm a}$ \gro\/ Viewing Period. \\
$^{\rm b}$ {Scaled to a single OSSE detector.} \\
$^{\rm c}$ {For 50--300 keV in units of $10^{-10}$ erg cm$^{-2}$ s$^{-1}$.} \\
$^{\rm d}$ {Pointed \xte\/ observations with Greek letters giving their
variability in the classification of Belloni et al.\ (2000).}         \\
$^{\rm e}$ {The {\it RXTE}/ASM 2--12 keV flux.} \\
$^{\rm f}$ {These errors give the standard deviation of the flux
variability.}
}

\footnotesize
\begin{center}
{\sc TABLE 2\\ Model parameters of the broad-band spectra}
\vskip 2pt
\begin{tabular}{ccccccccccccc}
\hline
\hline
&\\
OSSE & $\nh$ & $kT_{\rm bb}$ & $\lh/\ls$  & $\lnth/\lh$ & $\Gamma_{\rm inj}$ &
$\tau_{\rm i}$ & $\tau$\tablenotemark{a} & $kT$\tablenotemark{a} & $\Omega/ 2
\pi$ & $F_{\rm bol}$\tablenotemark{b} & $L_{\rm
iso}$\tablenotemark{c} & $\chi^2/\nu$\tablenotemark{d} \\

{VP} & [$10^{22}$ cm$^{-2}$] & [keV] & & & & & & [keV] & & & \\
&\\
\hline
&\\
619 & $3.2^{+0.3}_{-0.4}$ & $1.37^{+0.05}_{-0.05}$ & $0.48^{+0.03}_{-0.03}$ &
$0.23^{+0.31}_{-0.13}$ & $3.0^{+0.8}_{-0.6}$ & $0.26^{+0.34}_{-0.26}$ & 0.39 &
66 & $0.96^{+0.50}_{-0.26}$ & 3.5 & 6.5 & 91/130  \\
813 & $6.1^{+0.5}_{-0.4}$ & $1.35^{+0.04}_{-0.02}$ & $0.29^{+0.02}_{-0.01}$ &
$0.12^{+0.02}_{-0.01}$ & $2.2^{+0.2}_{-0.2}$ & $4.42^{+0.14}_{-0.06}$ & 4.43 &
3.6 & 0 (fixed) & 8.9 & 17 & 60/90 \\
\hline
\end{tabular}
\label{t:fits}
\end{center}
\setcounter{table}{2}
{
$^{\rm a}$ The total optical depth and the electron temperature calculated from
energy and pair balance for the best-fit model (i.e., not free parameters).\\
$^{\rm b}$ The unabsorbed bolometric flux of the model in units of
$10^{-8}$\,erg\,cm$^{-2}$\,s$^{-1}$.\\
$^{\rm c}$ The model luminosity assuming isotropy and a distance of 12.5 kpc in
units of $10^{38}$\,erg\,s$^{-1}$.\\
$^{\rm d}$ The values of $\chi^2/\nu<1$ result here from the assumed systematic
errors, and are not due to overmodeling.
}

\end{document}